# Shallow acceptor state in ZnO realized by simple irradiation and annealing route

S. Pal and D. Jana

*Department of Physics, University of Calcutta, 92 Acharya Prafulla Chandra Road, Kolkata 700009, India*

T. Rakshit

*Department of Physics, Indian Institute of Technology, Kharagpur 721302, India*

S. S. Singha

*Department of Physics, Bose Institute, 93/1, Acharya Prafulla Chandra Road, Kolkata 700009, India*

K. Asokan

*Inter University Accelerator Centre, Post Box 10502, Aruna Asaf Ali Marg, New Delhi 110067, India*

S. Dutta

*[d]Department of Physics, Rammohan College, 102/1, Raja Rammohan Sarani, Kolkata 700009, India.*

A. Sarkar[*]

*Department of Physics, Bangabasi Morning College, 19 Rajkumar Chakraborty Sarani, Kolkata 700009, India*

[*]Corresponding author: Tel. ++ 91 33 2360 7586, e-mail- sarkarcal@gmail.com

**Abstract**

Activation of shallow acceptor state has been observed in ion irradiated and subsequently air annealed polycrystalline ZnO material. Low temperature photoluminescence (PL) spectrum of the sample exhibits clear signature of acceptor bound exciton (ABX) emission at 3.360 eV. The other two samples, pristine and ion irradiated (without annealing), however, do not show acceptor related PL emission in the nearby energy region. Electron transition from shallow donor (most probable site is interstitial zinc for undoped ZnO) to such newly formed shallow acceptor state creates new donor-acceptor pair (DAP) luminescence peak ~ 3.229 eV. ABX and DAP peak energy positions confirm that the acceptor is N related. The acceptor exciton binding energy has been estimated to be $180 \pm 15$ meV which is in conformity with earlier reports. The activation of shallow acceptors without any source of atomic nitrogen can only be possible through diffusion of molecular nitrogen inside the sample during annealing. The $N_2$ molecules get trapped at bulk defect sites incorporated by ion irradiation and subsequent annealing. X-ray diffraction (XRD) and Raman spectroscopic (RS) investigation have been employed to probe the changing defective nature of the ZnO samples. Irradiation induced increased disorder has been detected (both by XRD and RS) which is partially removed/modified by annealing at 300 $^o$C. Simultaneous activation of molecular nitrogen acceptor in purposefully defective ZnO is the key finding of this work. Results presented here provide a simple but controlled way of producing shallow acceptor state in ZnO. If optimized through suitable choice of ion, its energy and fluence as well as the annealing temperature, this methodology can trigger further scope to fabricate devices using ZnO epitaxial thin films or nanowires.

Nitrogen being the prime suspect of inducing p-type conductivity in traditionally n-type ZnO, large effort has been put forward to dope ZnO by N using various techniques [1-15]. Obviously, N doping replacing an O atom is easier as the ionic radius of N atom is closest to that of O atoms compared to other group V elements. Back in 1988, Gutowski et al. experimentally [16] pointed out the presence of shallow acceptor complexes in ZnO. Further, successful efforts were made by Look et al. [2], Zeuner et al. [3] and Tsukazaki et al. [4] to fabricate p-type ZnO. A brief summary of other studies can be found in ref. 1. However, optimization of defects, chemical nature of the incorporated N atoms and reproducibility of such results have remained as major source of concern till date [17]. Theoretical calculation [9] followed by its experimental [10] confirmation have revealed that N in an O vacancy site ($N_O$) is indeed an acceptor but its energy level (~ 1.3 eV above the valence band maximum (VBM)) does not allow sufficient number of holes to be activated at room temperature (corresponding thermal energy ~ 25 meV). So the shallow N related acceptor as reported by several research groups are most probably not $N_O$. It has also been pointed out that in ZnO, Zn rich configuration cannot be made p type by N doping only [5,9]. On the other hand, it is natural that abundant Zn vacancies ($V_{Zn}$) are not helpful to achieve the hole induced conductivity as $V_{Zn}$ is also a deep acceptor in ZnO [18]. So the focus has been shifted towards the generation of defect complexes involving N and $V_{Zn}$ with energy levels sufficiently close (within few hundred meVs) to the VBM. It has been reported [19] that $V_{Zn}$–$N_O$–$H^+$ type stable acceptors with hole binding energy ~ 130 meV in requisite concentration can be formed in ZnO. Recent first-principles calculations [17] also emphasizes the presence of H to stabilize $V_{Zn}$–$N_O$ related shallow acceptors. However, the formation of such complexes in ZnO is rather tricky and there exists an alternative possibility of forming $V_O$–$N_{Zn}$ (O vacancy-N at Zn site) pair defects which are donors in ZnO [19]. In another track, acceptor doping in ZnO using $N_2$ or $NH_3$ molecules have also been explored [20] and recently in a more convincing way [21,22]. However, debates and search for new avenues of purposefully controlled N doping are continuing till date [23]. In this article, we have provided clear cut evidence of evolution of shallow acceptor state in ZnO through ion irradiation and annealing route. It has been understood that the most probable nature of this shallow acceptor is molecular nitrogen at $V_{Zn}$ and/or Zn-O di-vacancy ($V_{ZnO}$) sites.

Commercial polycrystalline ZnO powder (99.99%, Sigma-Aldrich, Germany) have been pelletized and then annealed at 500 ºC for four hours and cooled slowly (30 ºC/30 min). The pellets have been pre-annealed before irradiation to make the sample free from any residual organic materials or adsorbed species like $H_2O$ or $H_2$ if any [24]. The pre-annealed pellet has been irradiated with 96 MeV O ions with fluence $2.3 \times 10^{13}$ ions/cm$^2$ at Inter university accelerator centre (IUAC), New Delhi, India. Choice of such fluence ensures, but not much, special overlapping of collision cascades of the projectiles, taking each of their lateral straggling ~ 200 Å (estimated using SRIM software [25]). The irradiated sample has been divided into two parts. One of them has been annealed at 300 ºC for four hours followed by furnace cooling. These samples are hereafter referred as: ZnO-U (unirradiated), ZnO-I (irradiated) and ZnO-IA (irradiated and annealed). X-ray diffraction (XRD) patterns have been recorded using powder x-ray diffractometer (model: X'Pert Powder, PANalytical) using Cu $K_α$ (1.54 Å) radiation. The diffraction pattern of all the samples have been refined by Rietveld analyses [26] using the MAUD software [27]. According to Warren's treatment of fault probability analysis [28] three kinds of planar defects, namely intrinsic (*α′*) and extrinsic (*α′′*) deformation, and twin (*β*) fault probabilities have been used as refinable parameters. Photoluminescence (PL) spectra in the range (10-300 K) have been recorded using 325 nm He-Cd laser as excitation source (output power 45 mW) and a TRIAX 320 monochromator fitted with cooled Hamamatsu R928 photomultiplier detector. Taking the absorption coefficient ($\alpha_{abs}$) of ZnO to be $1.6 \times 10^5$ cm$^{-1}$ at 325 nm [29], the characteristic penetration depth ($1/\alpha_{abs}$) of the 325 nm laser excitation can be estimated ~ 60 nm. So, the observed PL emission spectra will contain information about the region up to few tens of nanometer from the upper surface of the sample. The Raman measurements were performed using a micro Raman set-up consisting of a spectrometer (Lab RAM HR Jovin Yvon) and a Peltier cold CCD detector. A He-Ne laser with wavelength of 633 nm was used as an excitation light source. Sheet resistance have been measured by usual two probe method (with deposited gold contacts) using Keithley K2000 multimeter and K2400 current sources.

The maximum penetration depth of 96 MeV O ion beam in ZnO is about 59 μm as estimated by SRIM [25] simulation software. In this calculation, the density of polycrystalline ZnO has been taken as 4.5 gm/cm$^3$ and the displacement threshold energies of Zn and O atoms are 34 eV and 44 eV respectively [30]. Along the path of the ion inside target material, it loses energy by both inelastic and elastic collisions known as electronic energy loss ($S_e$) and nuclear energy loss ($S_n$) respectively. The knockout of target atoms from their lattice positions takes place due to $S_n$ only. $S_e$ induces excitation and ionization of target electrons and thereby causing defect re-organization to some extent. Figure 1 shows the variations of $S_e$ and $S_n$ with penetration depth calculated using SRIM. Within the first 100 nm from the surface, from where most of the PL signal comes, the values of $S_e$ and $S_n$ are 117.1 eV/Å and 0.068 eV/Å respectively. So, $S_e$ is about 1730 times larger than that of $S_n$, which is quite different to the situation for low or medium energy ion irradiation on ZnO [31]. SRIM calculation also indicates that one 96 MeV O ion creates ~ 2400 vacancies ($V_{Zn}$ : $V_O$ ~ 2:1) most of them are populated at the last 5 μm of the range of the projectile. It is well known [32,33] that dynamic recovery of irradiation generated defects is very high in ZnO. Considering 99 % of the $V_{Zn}$s get recovered immediately after generation and $V_O$s are more or less stable [32], one can estimate roughly one $V_{Zn}$ and fifty $V_O$s in $10^7$ atoms within the subsurface 100 nm region of ZnO-I sample. Annealing at 300 °C should induce at least four fold modifications in the ZnO-I sample, recovery of almost all the unstable interstitial defects, recovery of a fraction (unknown) of $V_{Zn}$s [33-35], migration and agglomeration of defects in the form of $V_{Zn}$ clusters and $V_{ZnO}$ [34] and starting of generation of $V_O$s [36,37].

Low temperature PL spectra (10-300 K) have been recorded of all the three samples. Figure 2 (upper panel) shows 10 K near band edge (NBE) PL spectra of the three ZnO samples. The donor bound exciton (DBX) emission has been found at the same position (3.365 eV) for all the three samples which indicates that same donors are present in all three samples. Based on previous literature, the DBX has been assigned as interstitial zinc ($I_{Zn}$) related emission [31,38]. Besides DBX, another intense but broad peak has been noticed around 3.313 eV in all 10 K PL spectra. This peak contributes up to room temperature (RT) PL spectrum as is seen in figure 2 (lower panel). The intensity of this peak is highest in the ZnO-U sample spectrum. Electron transition from conduction band (CB) to the neutral acceptor level (free to bound, FB transition) can be the origin of that peak [39]. Acceptor energy ($E_a$) of the related acceptor level can be calculated using the formula: $E_a = E_g(0) - E(FB)$ where $E_g(0) = 3.437$ eV, band gap of wurtzite ZnO bulk at 0 K and $E(FB)$ is the PL peak emission energy (here, at 10 K) of the FB transition. Here, $E_a$ is found to be 124 ± 5 meV which is in close agreement [39-42] with the values found earlier (using different methods [39]) for the 3.313 eV PL peak. This particular acceptor has been assigned as a complex of $V_{Zn}$, which populates near the crystal imperfections and gets reduced due to high temperature annealing [31,41,42]. It is interesting to note that irradiation also causes a decrease of the intensity of this PL emission particularly compared to that of the DBX emission. A comparative picture (Figure 3), on this regard, has been plotted to show the effect of different ion species. Energy deposition by the impinging ions enhances diffusion and displacement of atomic elements of the target. The most probable [32] recovery site for ZnO is $V_{Zn}$ and thereby a reduction of 3.313 eV PL takes place. To some extent this is similar to high temperature annealing where thermally induced $V_{Zn}$ migration and recovery [31,41] takes place. A closer look to the 10 K PL spectrum of the ZnO-IA sample reveals a noticeable broadening of the DBX emission. One might think that this is due to the enhancement of shallow donor concentration due to 300 ºC annealing [43]. Of course, DBX intensity is also increased for ZnO-IA sample, however, the DBX broadening is not solely due to donors (discussed later). The RTPL NBE peak is also broadened for ZnO-IA sample compared to that of the ZnO-U sample (Figure 2 inset). A pronounced negative thermal quenching (NTQ) of PL intensity near RT in the photon energy range 2.42- 3.28 eV has also been seen for the ZnO-I sample (Figure 4, upper panel). This is a clear manifestation of shallow carrier traps, just above the valence band (VB), created by 96 MeV O ion irradiation. Indeed, NTQ of PL near RT is not commonly reported and is characteristically different from such phenomenon below 100 K. NTQ of PL below 100 K results from a competition [44] between two or more shallow donors to trap/release electrons as the temperature increases from 0 K. In fact, O ion irradiation (700 keV) on ZnO single crystal also bears similar features near RT (Figure 4 inset). Specific role of energetic O ions to stabilize shallow (as well as deep) acceptor states in ZnO needs further investigation. Increased weight around 3.16 eV, 3.06 eV, 2.8 eV, 2.4 eV and 2.2 eV in the RTPL spectrum has been observed after irradiation (sample ZnO-I). It is difficult to assign 3.16 eV and 3.06 eV PL emission at RT to particular defect sites. However, both of these PL peaks are either isolated $V_{Zn}$ related or $O_{Zn}$ type, as envisaged theoretically and experimentally [31,45,46]. Broad PL peak ~ 2.8 eV can be a signature of transition from different ionization states of Zn interstitials to VB band or shallow acceptor levels [47,48]. In this work, enhanced 2.8 eV PL emission is correlated with lower sheet resistance (nearly $1/3^{rd}$ of ZnO-U) for the ZnO-I sample. Disordered ZnO lattice with abundant under-coordinated Zn atoms can give rise to lower resistance of the sample [31]. The 2.4 eV PL emission is generally attributed to singly ionized $V_O$ centers [49]. The PL emission ~ 2.2 eV is reported to originate [48] from doubly ionized $V_O$ transformed from a singly ionized one via capturing hole in the grain boundary depletion region. Without ruling out such possibility, we want to note that small yet significant increase in PL emission ~ 2.2 eV takes place after annealing (ZnO-IA sample). This may be due to formation of $V_{ZnO}$ type defects as the predicted [18] transition energy of which is ~ 2.2 eV. Annealing affects the broad 2.8 eV and 3.06 eV emissions to drop drastically. It can be understood that 300 °C annealing is enough to recover related defect centers. In fact, the (002) peak of XRD (not shown) for ZnO-IA sample is positioned at $2\theta = 34.749°$ in comparison to $2\theta = 34.593°$ for ZnO-U and $2\theta = 34.606°$ for ZnO-I. Shift of (002) XRD peak to higher angle (compared to ZnO-I) can be due to partial recovery of interstitial as well as vacancy defects [50]. The changing nature of defective state from ZnO-U to ZnO-IA samples have also been manifested in the Rietveld analyses of the x-ray diffractograms as we shall mention later.

In the light of the above results, a comparative study on the NBE PL spectra of the three samples can be discussed. Closer look to the 20 K PL spectrum of ZnO-IA indicates a signature of shallow acceptor bound exciton (ABX) at 3.360 eV (Figure 5). As the temperature increases more and more electrons move to the CB and ABX emission becomes more facile compared to the DBX one. The ABX emission intensity surpasses that of the DBX at 40 K (Figure 5) and remains visible up to RT (Figure 4 lower panel). Indirectly, presence of ABX is re-confirmed from the fact that free exciton (FX) is not at all visible for ZnO-IA sample. On the other hand, FX emission can be distinctively identified from 30K in the PL spectra of ZnO-U and ZnO-I samples (Figs. 2 and 4 respectively). Further confirmation of the presence of shallow acceptors in the ZnO-IA sample comes from the presence of donor acceptor pair (DAP) transition at 3.229 eV. In fact, the nature of PL spectrum for this sample in the energy range 3.20 to 3.27 eV is clearly different from those of the other two samples (Figure 2 upper panel). Earlier, DAP emission from shallow donors to shallow N related acceptor has been found at 3.235 eV [3], 3.238eV [2,51] and 3.241 [13]. Broad emission ~ 3.23 eV has also been found for N and O co-implanted ZnO single crystal [24]. Previous studies have indicated Zn excess in the sample can hinder the activation of shallow acceptors [5,9]. O ion irradiation and subsequent annealing in the present study efficiently creates $V_{Zn}$ and $V_{ZnO}$ type defects. Annealing reduces different energy states related to $I_{Zn}$ defects ~ 2.8 eV. Annealing also recovers a fraction (unknown) of $V_{Zn}$ defects but also promotes $V_{ZnO}$ formation. One cannot rule out formation of $N_{Zn}$-$2V_{Zn}$ shallow acceptor complex [12] during annealing. However, in these samples without any source of atomic N, $N_{Zn}$-$2V_{Zn}$ defects are unlikely to be formed. Also the nature and abundance of defects species generated due to irradiation (as estimated by SRIM) does not favour the formation of $N_{Zn}$-$2V_{Zn}$ type defects.

Using the DAP energy position, one can calculate the acceptor binding energy ($E_A$) using the well known formula

$$E_A = E_g(0) - E_D - (E_{DAP} - \gamma N_D^{1/3})$$

Where $E_D$, $E_{DAP}$ and $N_D$ denote the donor binding energy, DAP emission energy position and concentration of donor defects respectively. For ZnO, $E_g(0) = 3.437$ eV and $E_D = 0.043$ eV can be used [28] for calculation. $\gamma$ has constant value of $3.0 \times 10^{-5}$ meV-cm. If one assumes that $N_D$ for these present samples are within the order of $10^{16}$-$10^{18}$/cm$^3$ (good approximation for non-degenerate semiconductors), the values of $E_A$ comes to be $180 \pm 15$ meV. Such an estimation nicely agrees with findings of different groups [1,3,22] for N related acceptor in ZnO as well as with most recent results [14]. Further confirmation comes from the thermal quenching (Figure 6) of ABX peak intensity of ZnO-IA sample when fitted with the following equation

$$I(T) = I(0)/[1 + A\exp(-E_b/k_BT)]$$

where, $I(T)$ is the intensity at any temperature $T$, $E_b$ is the activation energy of the thermal quenching process and $k_B$ is Boltzmann's constant. Here $E_b$ comes out to be $16.6 \pm 0.9$ meV, which is very close to the localization energy ($E_{loc} = 15.9$ meV) of the acceptor indicating it as a neutral one. The relationship of $E_{loc}$ and $E_D$ is well established for donors in ZnO (celebrated Hayne's rule), however, it is not clearly known for acceptors. Existing literature [52] provides us a most probable ratio of $E_{loc}/E_A \sim 0.1$. So a rough estimate of $E_A \sim 159$ meV in the present study is in close agreement with $180 \pm 15$ meV found from the DAP energy position. Thermal quenching of DBX peak intensities for ZnO-U and ZnO-I samples have also been fitted with the same equation and $E_b$ values comes out to be $8.5 \pm 0.7$ meV and $8.1 \pm 0.6$ meV respectively. Combining the peak energy position (DBX at 3.365 eV) and $E_b$ values, it is once again verified that no new type of shallow donors have been generated due to irradiation and/or annealing at 300 ºC.

Børseth et al have found [53] the evolution of $V_{Zn}$ clusters in low energy N irradiated ZnO single crystal wafers. Such vacancy clusters grow in size after annealing at 600 °C in air. The authors, however, do not rule out the presence of $V_O$s inside such clusters. In fact, earlier studies using positron annihilation spectroscopy (PAS) clearly reveal the presence of cation-anion di-vacancy clusters in annealed [54] polycrystalline ZnO, ion irradiated ZnO and $TiO_2$ (refs. 55, 56) and irradiated plus annealed ZnO (ref. 57). Deep level transient spectroscopic investigation on Zn ion irradiated ZnO samples has revealed [58] the presence of deep traps, presumably vacancy clusters, but not agglomeration of single type vacancies. Recent theoretical simulations [59] based on kinetic perspective of native defects during annealing, have also supported this contention. Once formed, such clusters, particularly $V_{ZnO}$ types are quite stable [57] and dissociate above 700 °C. However, the growth in size of the vacancy clusters either during equilibrium (annealing) or non-equilibrium (ion irradiation) conditions, the probability of their transformation into optically inactive centers also increases. Present approach takes note that low $S_n$ value of 96 MeV O ions in ZnO can be advantageous in the sense that vacancies may not grow much during collision cascades. Subsequent annealing at 300 °C will help migration [60] of point defects to form stable $V_{ZnO}$ type divacancies. It is also expected that molecular $N_2$ diffusion [13] in the sample during annealing in air is possible and a fraction of $V_{ZnO}$ can be filled thereby. In fact, unintentional [15] and intentional [61] $N_2$ incorporation is common in defective ZnO material. This is a simple method for acceptor doping in ZnO and further theoretical and experimental studies should consider the possibility of molecular nitrogen both at $V_{Zn}$ and $V_{ZnO}$ defects as well as slightly larger clusters [59] like $2V_{Zn}-V_O$. The point we note that choice of temperature and gaseous environment during annealing is very much important. The study [7] by Myers et al., where irradiation has been done at an elevated temperatures (say $T_I$), clearly shows that ion fluence and $T_I$ combination can dramatically change the accumulated disorder in ZnO. Also it is known that annealing in Ar gas atmosphere [22], even after N doping, does not lead to activation of N dopants, rather it may promote $V_O$ formation. Annealing at higher temperatures (700 °C) may de-activate acceptor dopants[6] or helps in the formation of $N_O$ type deep acceptor centers [62]. It can be conjectured that low temperature annealing [5,13] or growth [6] particularly below 450 °C [8] are effective to promote $N_2$ related acceptor state in ZnO and to avoid huge $V_O/I_{Zn}$ formation [36,37]. Also to note, grain boundaries (GB) in polycrystalline ZnO always contain clusters of $V_{ZnO}$ (or $2V_{Zn}-V_O$, $2V_O-V_{Zn}$ etc.) type defects detected [34] by PAS. $N_2$ incorporation at GB defects may not be useful as large voids are optically and electrically deleterious for the system. So, the presence of bulk isolated $V_{Zn}$ and/or $V_{ZnO}$ type defects in requisite concentration is the most important criteria. In this connection, single crystalline nanowires or epitaxial thin films will be more effective as a starting material.

To investigate the change in overall defective nature of ZnO due to O ion irradiation and subsequent annealing, Raman spectroscopy has been employed. The Raman spectra of ZnO-U, ZnO-I and ZnO-IA samples at RT have been shown in Figure 7. In spite of the claims made by various researchers [22,63], it is very difficult indeed to extract information about $N_2$ incorporation in ZnO from Raman spectrum. Particularly for ZnO-I spectrum, a broad background peaking ~ 1400 cm$^{-1}$ appears although the same is not seen for ZnO-U. This background is due to defect related luminescence from the irradiated sample. All the spectra have been taken by exciting the samples with 633 nm (1.96 eV) laser and the broad luminescence peak can be calculated ~ 1.75 eV. The excitation energy of the laser is much lower than the band gap (3.37 eV at RT) of ZnO, and hence band to band or shallow donor electron recombination are not associated with PL ~ 1.75 eV. Most probable sites responsible for such background PL are small size $V_{Zn}$ clusters [64]. After annealing, the overall background luminescence is reduced which indicates that a fraction of these defects are recovered or modified. Presence of characteristic Raman modes such as $E_2$ (low) at 100 cm$^{-1}$, $E_2$ (high) at 438 cm$^{-1}$ and even $A_1$(TO) at 380 cm$^{-1}$ of wurtzite ZnO structure indicates that major structural distortion has not been introduced due to irradiation (better viewed in right inset). Similar observation has also been reported by Myroniuk et al. after swift Xe ion irradiation [65] on ZnO films. However, damage has indeed been introduced due to high energy O ions, though small, as seen from the ratio of intensities of 438 cm$^{-1}$ and 582 cm$^{-1}$ Raman modes, highest for ZnO-U (7.7) and lowest for ZnO-I (1.8). The outcome of Rietveld analyses of the XRD spectra is in qualitative agreement with Raman spectroscopic results. The net deformation $\alpha$ (= $\alpha'$ - $\alpha''$) and the twin fault probability $\beta$ are increased for ZnO-I (2.02 × 10$^4$ and 3.79 × 10$^4$ respectively) compared to ZnO-U (0.21 × 10$^4$ and 0.06 × 10$^4$ respectively) and decreased subsequently on annealing (0.05 × 10$^4$ and 0.17 × 10$^4$ respectively for ZnO-IA). Finally, it is to be noted that Raman mode around 275 cm$^{-1}$ has not been found for any of the ZnO-U, ZnO-I and ZnO-IA samples. Such a peak, associated with either $N_O$ type defects or $N_O$ induced $I_{Zn}$ defects [66] is always seen in N irradiated ZnO and is shown in the left inset of Fig. 7.

In conclusion, prominent shallow acceptor bound exciton luminescence has been found in 96 MeV ion irradiated and subsequently air annealed polycrystalline ZnO. The ABX emission remains visible up to room temperature with acceptor binding energy 180 ± 15 meV. Results altogether indicate that molecular nitrogen chemically bound at small size $V_{Zn}$ and/or $V_{ZnO}$ defects are responsible for generating shallow acceptor levels close to the VBM. Similar shallow N related acceptors have been reported by several groups but, none to our knowledge, in this route. The methodology adopted in this study is simple with several controllable parameters for optimization such as, nature of starting ZnO material, ion species, energy and fluence along with annealing conditions. Present study provides basic data and understanding on this regard and further theoretical and experimental investigations would be encouraging.

**Acknowledgements**

Authors thank Prof. A. Singha, Department of Physics, Bose Institute, Kolkata for extending the facility of Raman measurements and helpful discussion regarding Raman data analysis. Authors also thank Dr. D. Sanyal, Variable Energy Cyclotron Centre, Kolkata for providing single crystals of ZnO sample and insights from his DFT calculation on similar ZnO system. Authors acknowledge helpful suggestions from Prof. S. K. Ray, Indian Institute of Technology, Kharagpur during PL data analysis. S. P. gratefully acknowledges University Grants Commission (UGC), Govt. of India for providing his RFSMS fellowship.

## Figure captions

FIG. 1. Estimated $S_e$, $S_n$ and vacancy concentration (in the inset) for 96 MeV O ions on ZnO target using SRIM software.

FIG. 2. NBE 10 K PL spectra of all the ZnO samples (upper panel) with corresponding RT PL spectra in the inset. Temperature evolution (10–300K) of PL spectrum for ZnO-U sample (lower panel).

FIG. 3. Variations of DBX intensity (3.365 eV) with respect to that of the FB emission (3.313 eV) with fluence of irradiation for different ion species.

FIG. 4. Temperature evolution (10–300 K) of PL spectrum for the ZnO-I (upper panel) and ZnO-IA (lower panel) samples. Clear NTQ of PL is seen at and above 250 K for ZnO-I sample. The inset shows similar NTQ of PL near RT for ZnO single crystal irradiated by 700 keV oxygen ions (fluence: $3\times10^{15}$ ions/cm$^2$).

FIG. 5. Temperature evolution (10–50K) of PL spectrum for ZnO-IA sample showing distinct ABX and DAP emissions at 3.360 eV and 3.229 eV respectively.

FIG. 6. Thermal quenching of DBX emissions (3.365 eV) for ZnO-U, ZnO-I and ABX emission (3.360 eV) for ZnO-IA samples.

FIG. 7. Raman spectra at RT for all the ZnO samples. Right inset shows the same spectra in a limited region. Prominent Raman modes for wurtzite ZnO have been indicated. The left inset shows evolution of new Raman mode at 275 cm$^{-1}$ due to N ion irradiation (energy: 625 keV, fluence: $1\times10^{16}$ ions/cm$^2$).

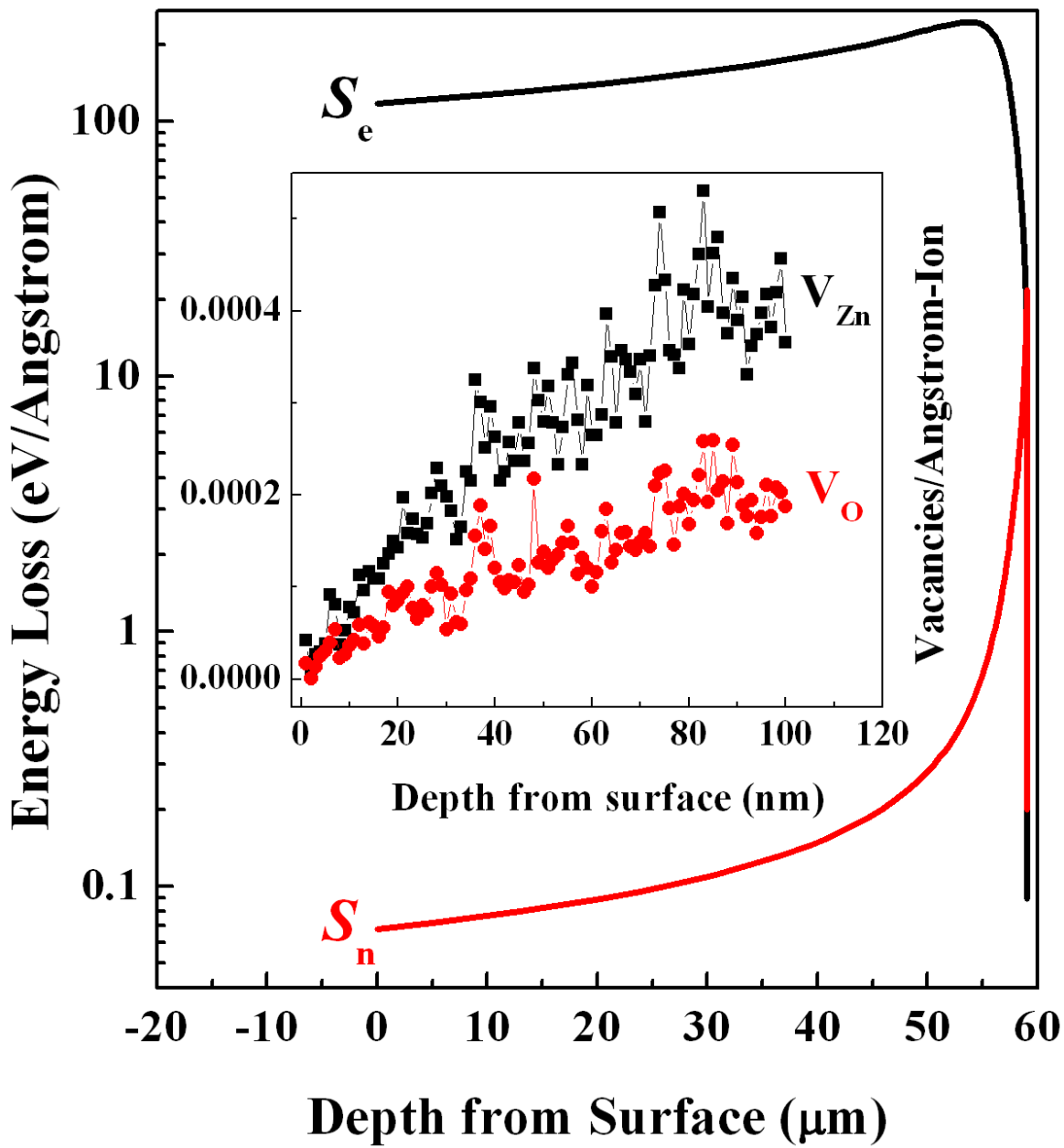

FIG. 1.

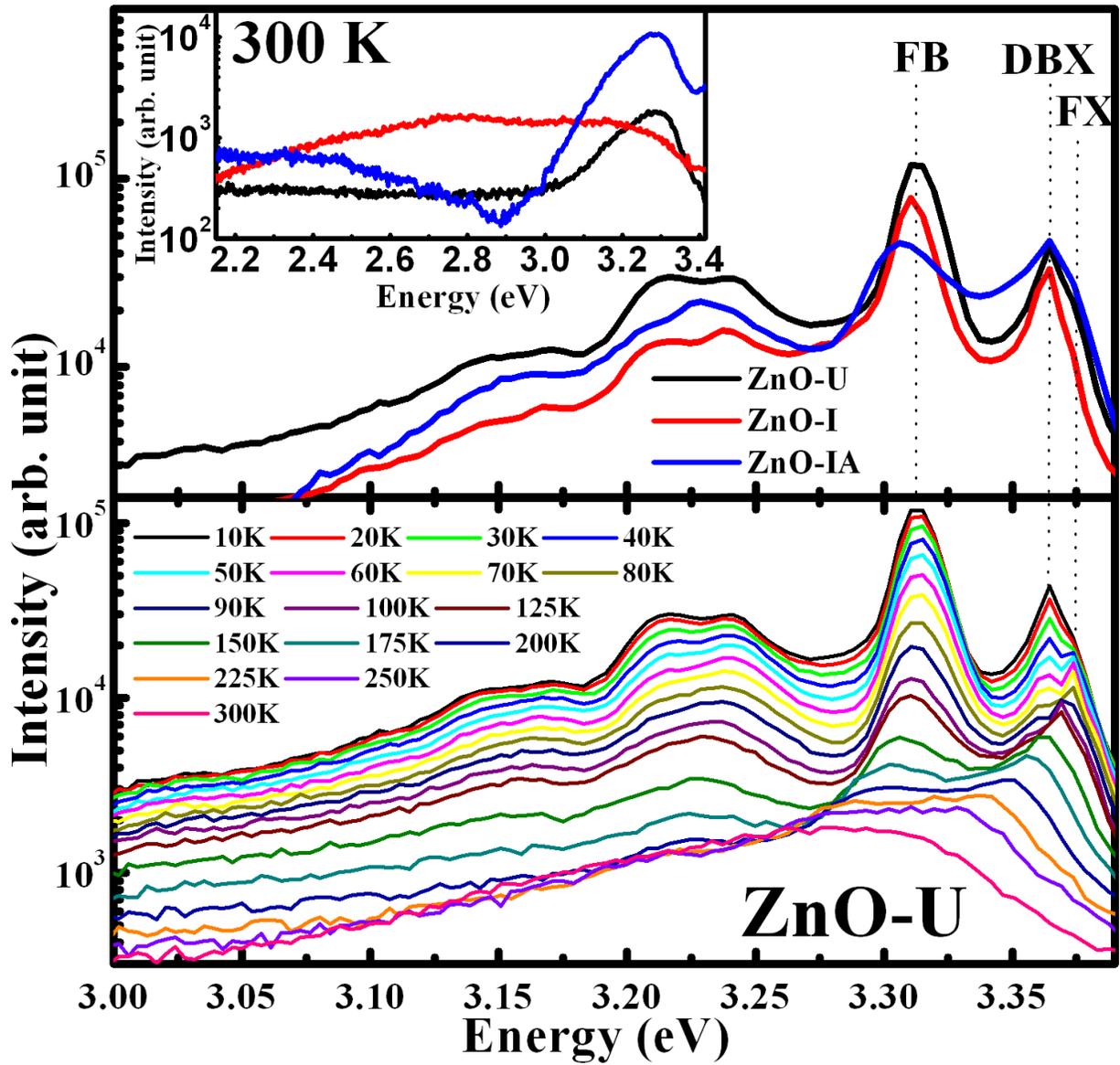

**FIG. 2.**

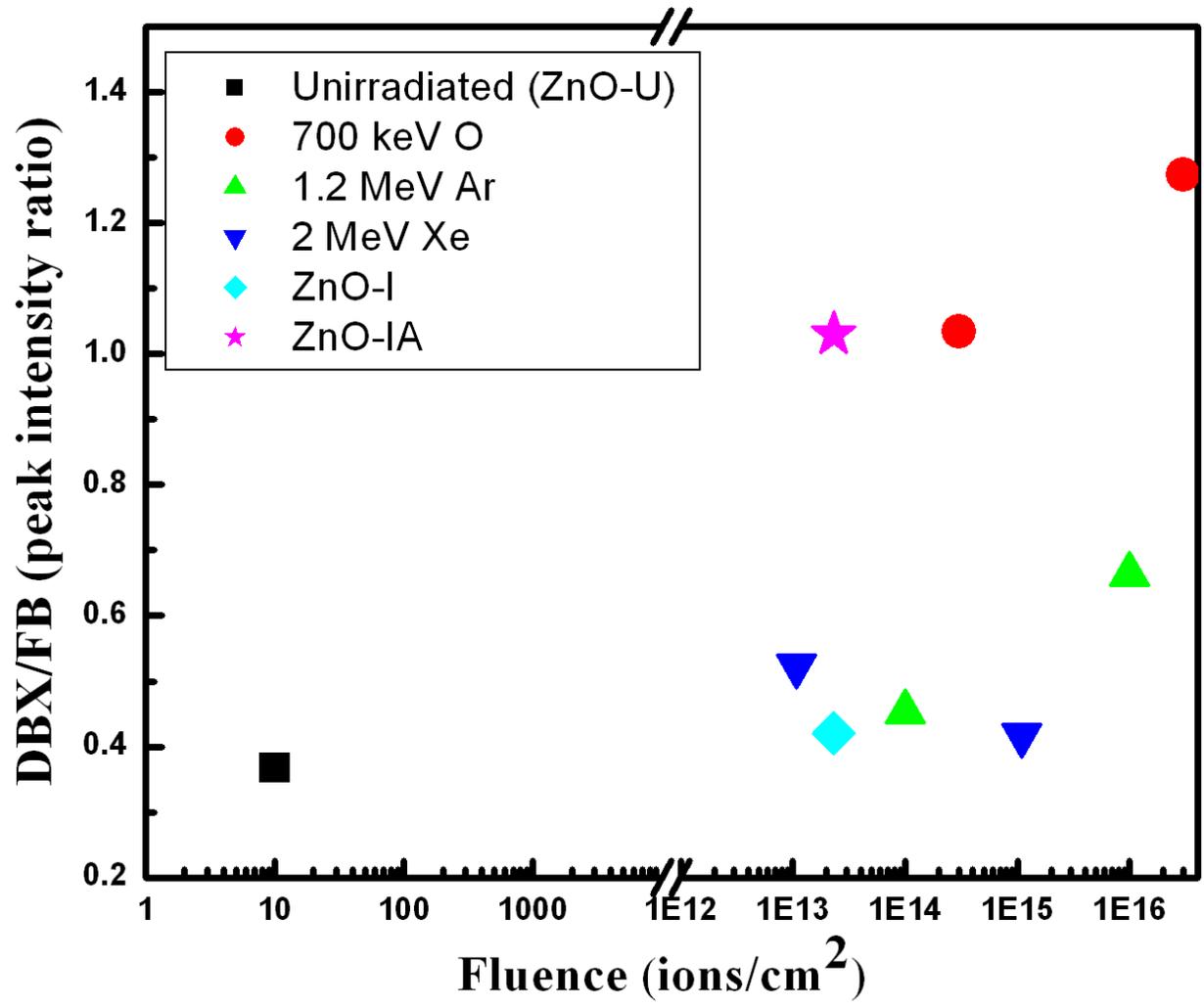

FIG. 3.

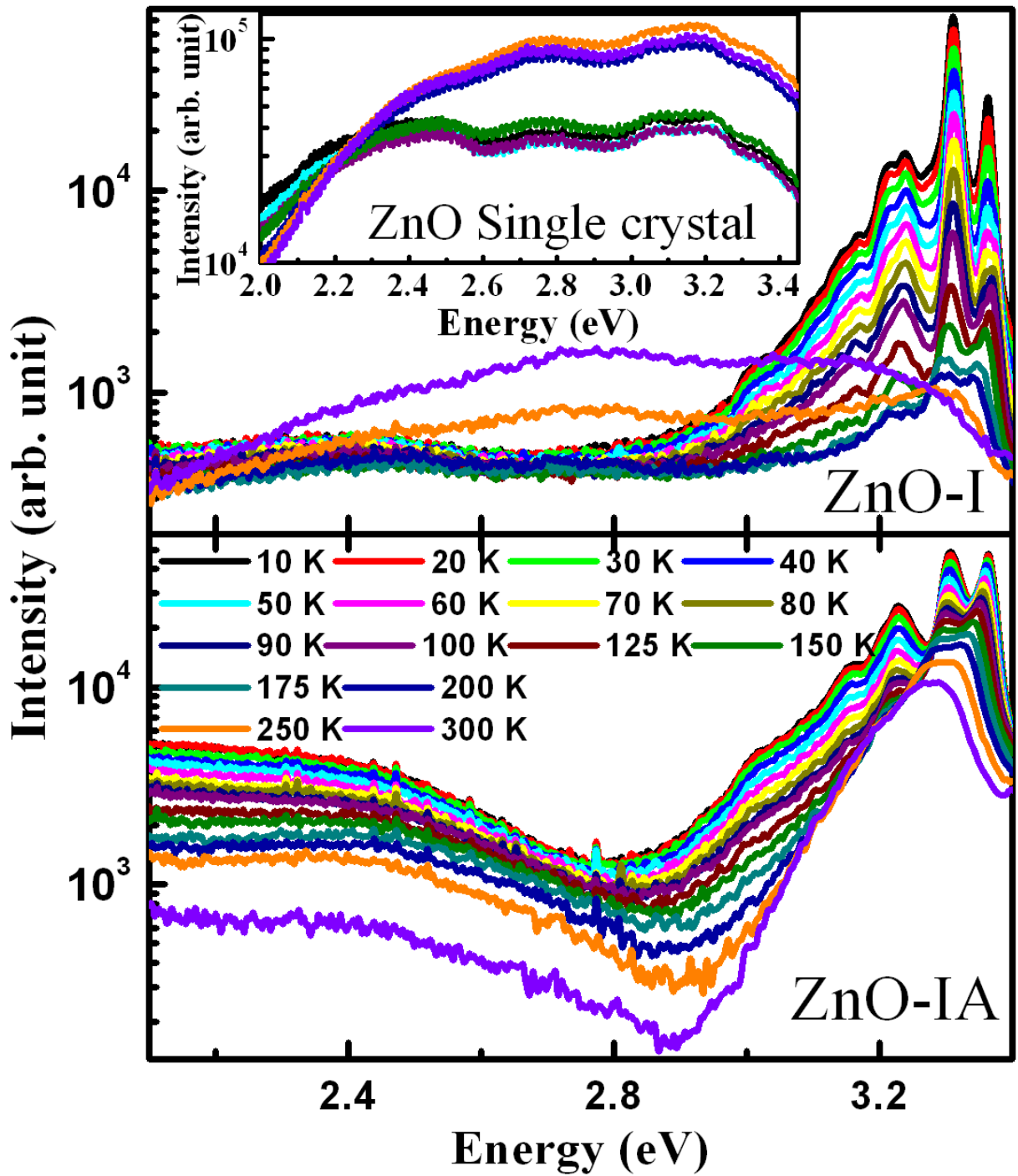

**FIG. 4.**

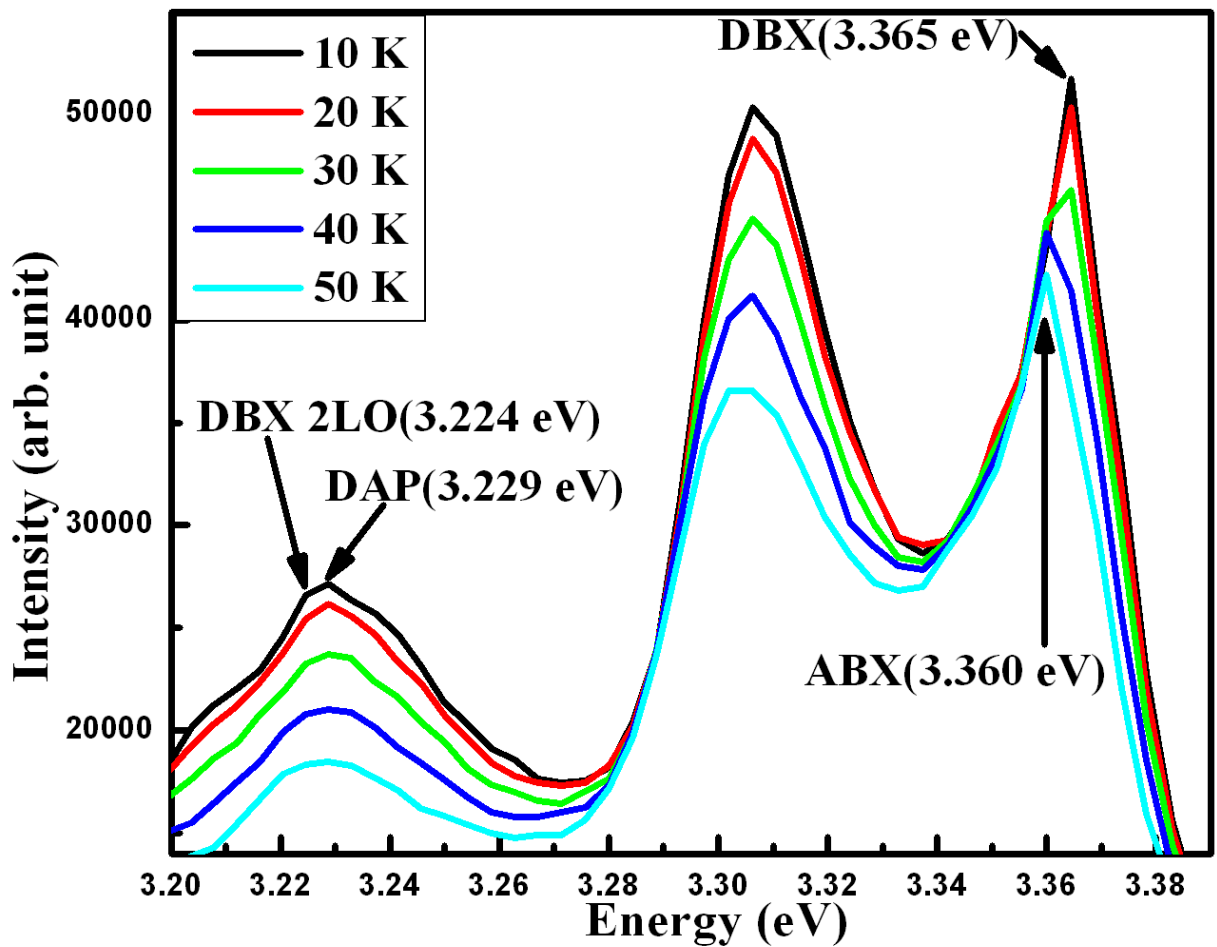

FIG. 5.

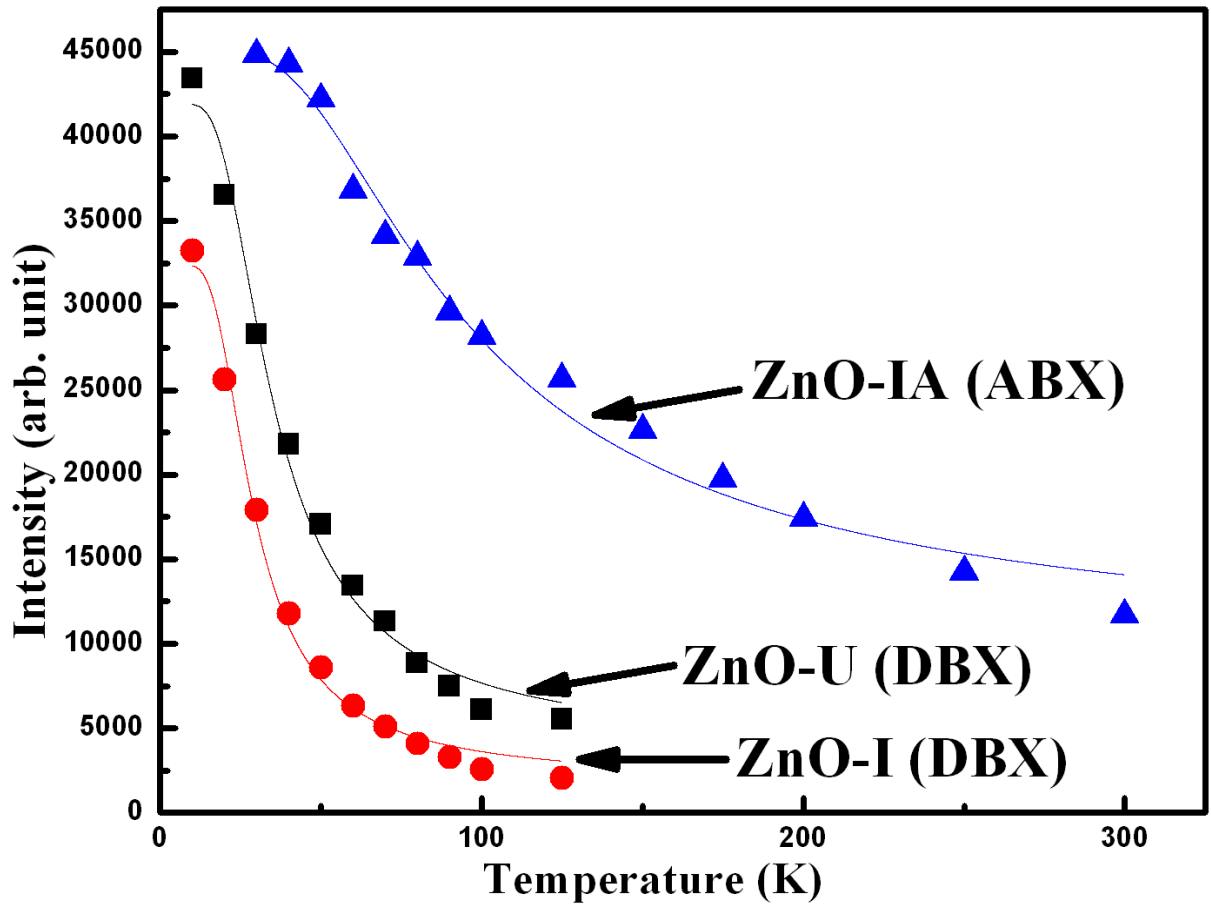

**FIG. 6.**

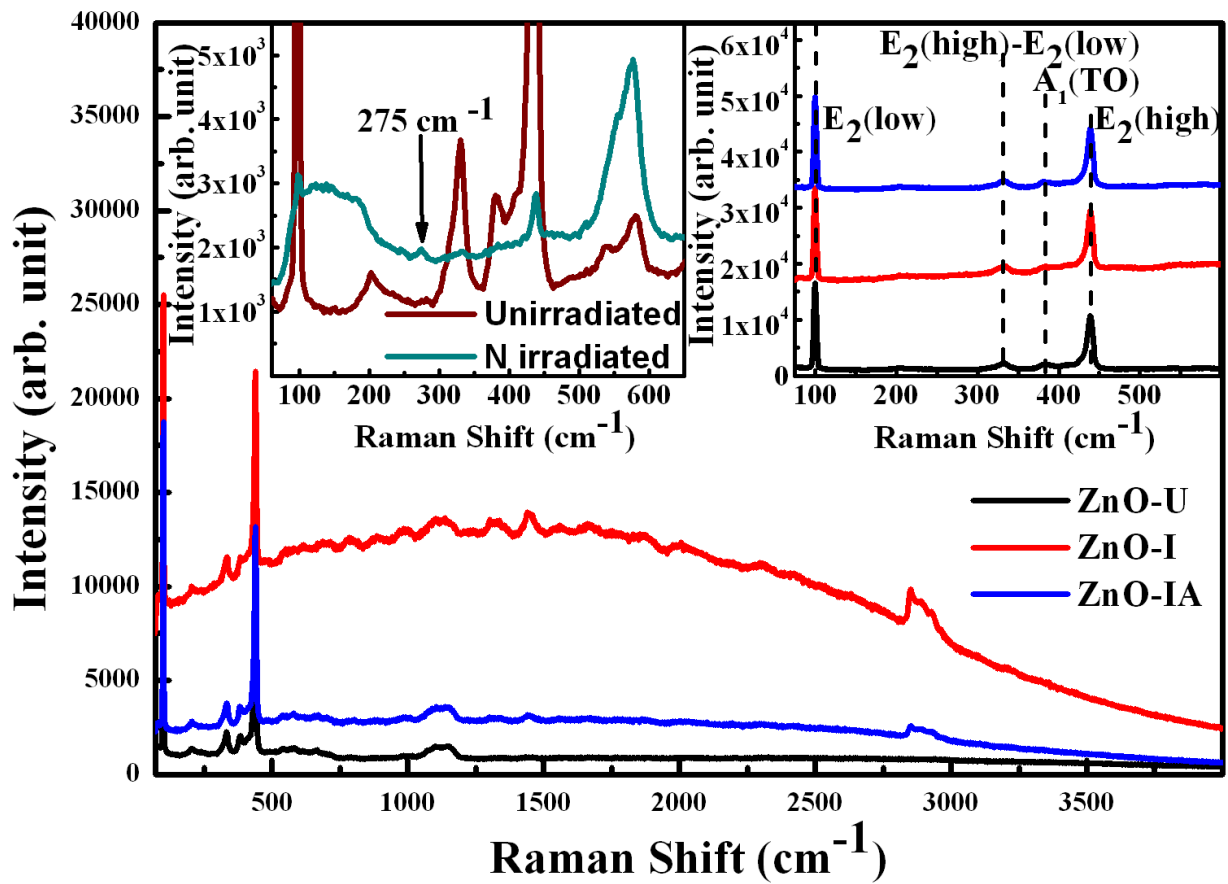

**FIG. 7.**